\documentclass[aps,prl,twocolumn]{revtex4-1} 
\usepackage{graphicx,subfigure}
\usepackage{amsmath,amssymb}
\usepackage{mathtools}
\usepackage{multirow}
\usepackage{textcomp}
\usepackage{float}
\usepackage{color}
\usepackage{ulem}
\usepackage{dsfont}
\usepackage{pdfpages}

\newcommand{\avg}[1]{\langle #1 \rangle}

\newcommand{\ket}[1]{\ensuremath{\left\vert{#1}\right\rangle}}

\newcommand{\uvec}[1]{\ensuremath{\hat{\mathbf{#1}}}}
\newcommand{\abs}[1]{\ensuremath{\left\vert{#1}\right\vert}}

\newcommand{\micro}[1]{\ensuremath{\mu\mathrm{#1}}}

\renewcommand{\micro}[1]{\ensuremath \mu\mathrm{#1}}

\renewcommand{\vec}[1]{\ensuremath{\mathbf{#1}}}

\newcommand{\f}{\ensuremath{h}}

\newcommand{\ddelta}{\ensuremath{\delta_c}}
\renewcommand{\f}{\ensuremath{f}}
\newcommand{\disp}{\mathcal{A}}

\makeatletter
\AtBeginDocument{\let\LS@rot\@undefined}
\makeatother

\begin{document}
\title{Photon-Mediated Spin-Exchange Dynamics of Spin-1 Atoms}
\author{Emily J. Davis, Gregory Bentsen, Lukas Homeier, Tracy Li, and Monika H. Schleier-Smith}
\affiliation{Department of Physics, Stanford University, Stanford, California 94305, USA}
\begin{abstract}
We report direct observations of photon-mediated spin-exchange interactions in an atomic ensemble.  Interactions extending over a distance of $500$~microns are generated within a cloud of cold rubidium atoms coupled to a single mode of light in an optical resonator.  We characterize the system via quench dynamics and imaging of the local magnetization, verifying the coherence of the interactions and demonstrating optical control of their strength and sign.  Furthermore, by initializing the spin-1 system in the $m_\f = 0$ Zeeman state, we observe correlated pair creation in the $m_\f = \pm 1$ states, a process analogous to spontaneous parametric down-conversion and to spin mixing in Bose-Einstein condensates.  Our work opens new opportunities in quantum simulation with long-range interactions and in entanglement-enhanced metrology.

\end{abstract}
\date{\today}


\maketitle

The hallmark of quantum information is its capacity to be non-local, encoded in correlations among entangled particles.  By contrast, the interactions between particles are necessarily local, restricting the quantum states that arise in nature.  Nevertheless, non-local interactions appear in a wide range of conceptual models, from holographic models of quantum gravity \cite{sachdev2015bekenstein} to spin models encoding hard optimization problems \cite{lucas2014ising,mertens1998phase} that are intimately connected to the physics of spin glasses \cite{strack2011dicke,gopalakrishnan2011frustration}.

Effectively non-local models can be generated in the laboratory by coupling atoms or solid-state qubits to optical or microwave resonators, where photons mediate long-range interactions \cite{majer2007coupling,van2013photon,leroux2010implementation,hosten2016quantum,norcia2018cavity,black2003observation,kollar2017supermode,leonard2017supersolid,landini2018formation,kroeze2018spinor}.  In atomic ensembles, interfacing photons with collective motional degrees of freedom has led to remarkable self-organization phenomena \cite{black2003observation,kollar2017supermode,leonard2017supersolid,landini2018formation,zhiqiang2017nonequilibrium,kroeze2018spinor} including supersolidity \cite{leonard2017supersolid}, while photon-mediated spin interactions \cite{andre2002coherent,sorensen2002entangling,schleier2010squeezing} have been harnessed to prepare squeezed states \cite{leroux2010implementation,hosten2016quantum} for quantum metrology.

Past experiments realizing cavity-mediated spin interactions have focused on manipulating and probing collective degrees of freedom \cite{leroux2010implementation,hosten2016quantum,norcia2018cavity,landini2018formation,zhiqiang2017nonequilibrium,kroeze2018spinor}.  For example, for atoms initialized in a spin-polarized state and uniformly coupled to a single cavity mode, the subsequent dynamics can be completely characterized by inferring moments of the total magnetization from measurements of the outgoing light.  In principle, photon-mediated interactions can also access richer many-body physics \cite{strack2011dicke,gopalakrishnan2011frustration,swingle2016measuring,colella2018quantum,mivehvar2017disorder}, including topological phases of matter \cite{hung2016quantum, mivehvar2017superradiant} and dynamical gauge fields \cite{zheng2016superradiance,ballantine2017meissner}.  However, fully benefiting from the non-local character of the interactions requires combining strong atom-light coupling with local addressing and imaging of spin dynamics.

Of particular interest for prospective applications in quantum simulation \cite{hung2016quantum,swingle2016measuring,colella2018quantum} are light-induced spin-exchange interactions.  Several theoretical proposals envision tuning the strength, sign, or spatial structure of spin-exchange couplings via optical drive fields \cite{hung2016quantum,swingle2016measuring,colella2018quantum}.  While spin-exchange interactions mediated by the vacuum field in a cavity \cite{hu2017vacuum,norcia2018cavity,lewis2018robust} have recently been detected \cite{norcia2018cavity}, achieving optical control over similar interactions requires a two-photon coupling between spin states, e.g., hyperfine or Zeeman states.  The latter encoding furthermore enables exploration of higher-spin models, including long-range-interacting analogs of spinor Bose condensates \cite{masson2017cavity,chang2005coherent,sadler2006spontaneous,klempt2010parametric,lucke2011twin,hamley2012spin,luo2017deterministic}.

In this Letter, we report direct observations of photon-mediated spin-exchange interactions in an ensemble of spin-1 atoms.  Spin excitations generated in one region of a spatially extended atomic cloud are observed to hop coherently over a distance of hundreds of microns.  We characterize the interactions via quench dynamics, demonstrating optical control of the interactions' strength and sign.  Furthermore, for a system initialized in the $m_\f=0$ Zeeman state, we observe light-mediated spin mixing, evidenced by correlated population growth in the $m_\f = \pm 1$ states.  An analog of spontaneous parametric down-conversion, this pair creation process paves the way to generating new many-atom entangled states.

\begin{figure}[htb]
\includegraphics[width=\columnwidth]{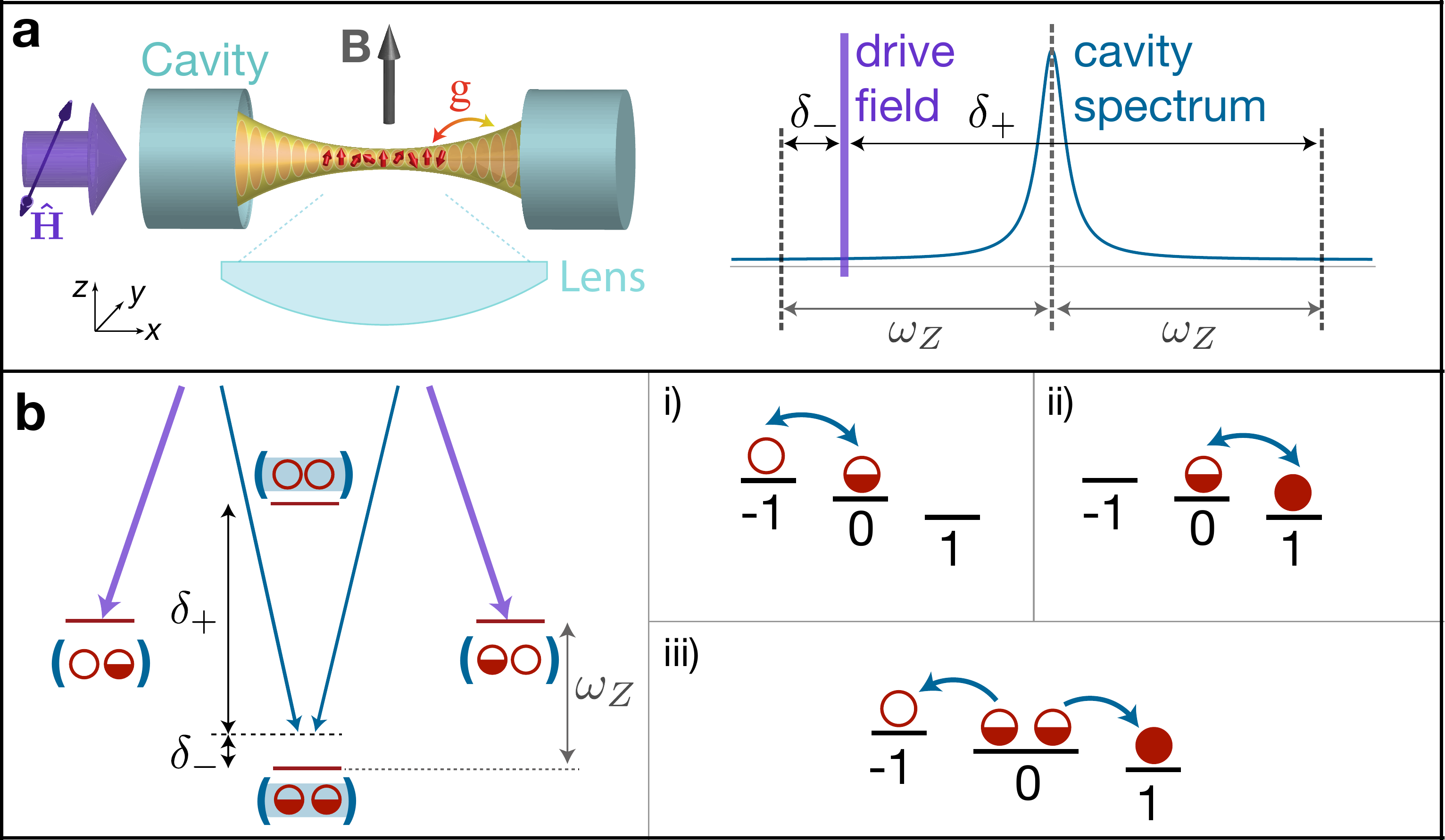}
\caption{\textbf{Experimental setup and scheme for generating spin-exchange interactions.}  (a) Driven cavity with atoms (red), transverse magnetic field, and imaging lens.  The drive field is detuned by $\delta_\pm$ from Raman resonances.  (b) Pairwise interactions are generated by one atom scattering a photon from the driven cavity mode (purple) into the orthogonally polarized cavity mode (blue), and a second atom rescattering the photon.  This mechanism can produce spin-exchange interactions (i-ii) or spin mixing (iii). Red circles indicate spin states $m=-1$ (empty), $m=0$ (half-filled) and $m=1$ (full).}\label{fig:exp_setup}
\end{figure}

The scheme for generating spin-exchange interactions is illustrated in Fig. \ref{fig:exp_setup}.  The building block is a Raman process in which an atom changes its internal state by absorbing a photon from a control field and emitting it into a cavity mode.  When the Raman coupling is resonant, its dominant effect is to induce superradiant decay \cite{kohler2018negative}.  For a control field detuned from Raman resonance, however, virtual emission into the cavity can induce a ``flip-flop'' process, wherein a second atom flips its spin by absorbing the virtual photon and rescattering it into the mode of the control field.

The flip-flop dynamics are described by an effective Hamiltonian \cite{SM}
\begin{equation}\label{eq:Hflipflop}
H = \hbar\sum_{i,j} \left(\chi_{ij}^+ f_i^+ f_j^- + \chi_{ij}^- f_i^- f_j^+\right),
\end{equation}
where $\mathbf{f}_i$ denote the spins of individual atoms, each pinned to a fixed location. 
The strengths of the spin-exchange couplings $\chi_{ij}^\pm$ are controlled by the amplitude of a drive field, as well as the spatial profile of the cavity mode.  The sign of the interactions is governed by the detunings $\delta_{\pm}$ from two Raman resonances, illustrated in Fig. \ref{fig:exp_setup}.  Hence, the interactions can be ferromagnetic or antiferromagnetic depending on the frequency of the control field.


To understand the dynamics that we expect to observe, we may view each spin-1 atom as a site that can hold up to two spin excitations.  The flip-flop process then corresponds to hopping of a spin excitation between two sites (Fig. 1b), mediated by converting the spin excitation into an intracavity photon.  Besides exchanging empty and singly occupied (or singly and doubly occupied) sites at arbitrary distances, this process can transform two singly occupied sites into a doublon-hole pair.  We will be able to observe either of these processes---spin exchange or pair creation---depending how we initialize the system.



We investigate the spin dynamics in a cloud of $N \sim 10^5$ rubidium-87 atoms trapped in a standing wave of 1560-nm light in a single-mode optical resonator.  The conduit for mediating interactions is a 780-nm cavity mode at large detuning $\Delta = -2\pi\times 10$~GHz from the $\ket{5S_{1/2}, \f=1}\rightarrow \ket{5P_{3/2}}$ transitions.  The coherence of the atom-light coupling at cavity center, where the mode has a 16-$\micro{m}$ waist, is parameterized by the single-atom cooperativity $\eta \equiv 4g^2/(\kappa\Gamma)=7.5$. Here, $2g = 2\pi\times 3.0(2)$~MHz is the vacuum Rabi frequency, $\Gamma = 2\pi\times 6$~MHz is the atomic excited-state linewidth, and $\kappa = 2\pi\times 200(50)$~kHz is the cavity linewidth.

The scheme for inducing spin-exchange interactions (Fig. \ref{fig:exp_setup}b) can be implemented either by directly driving the atoms or by driving the cavity.  We adopt the latter approach.  The atoms are initialized in the $\f=1$ hyperfine manifold, and we apply a uniform magnetic field $B \hat{z}$ transverse to the cavity axis (Fig. \ref{fig:exp_setup}a) to produce a Zeeman splitting $\omega_Z = \mu_B B/2$.  Spins placed in a superposition of Zeeman levels then undergo a Larmor precession that couples to the cavity via the Faraday effect, introducing a modulated birefringence.  For a cavity driven with horizontal ($\uvec{H}$) polarization, the atoms thus modulate the polarization of the intracavity field---or, equivalently, scatter photons from the $\uvec{H}$-polarized into the $\uvec{V}$-polarized cavity mode.  These scattered photons mediate the interactions among the spins.


To generate coherent interactions, we drive the cavity with a control field detuned from Raman resonance. Letting $\omega_c^N$ denote the cavity resonance frequency in the presence of the atoms, tuning the drive field to a frequency $\omega_d = \omega_c^N + \ddelta$ results in detunings $\delta_\pm = \ddelta \mp \omega_Z$ from the two Raman resonances shown in Fig. \ref{fig:exp_setup}.  While driving on either resonance ($\delta_\pm=0$) produces superradiant decay, with increasing detuning $\delta_\pm > \kappa$ we expect the decay to be suppressed relative to coherent interactions induced by the back-action of the cavity field on the atoms \cite{SM}.

\begin{figure}[htb]
\includegraphics[width=\columnwidth]{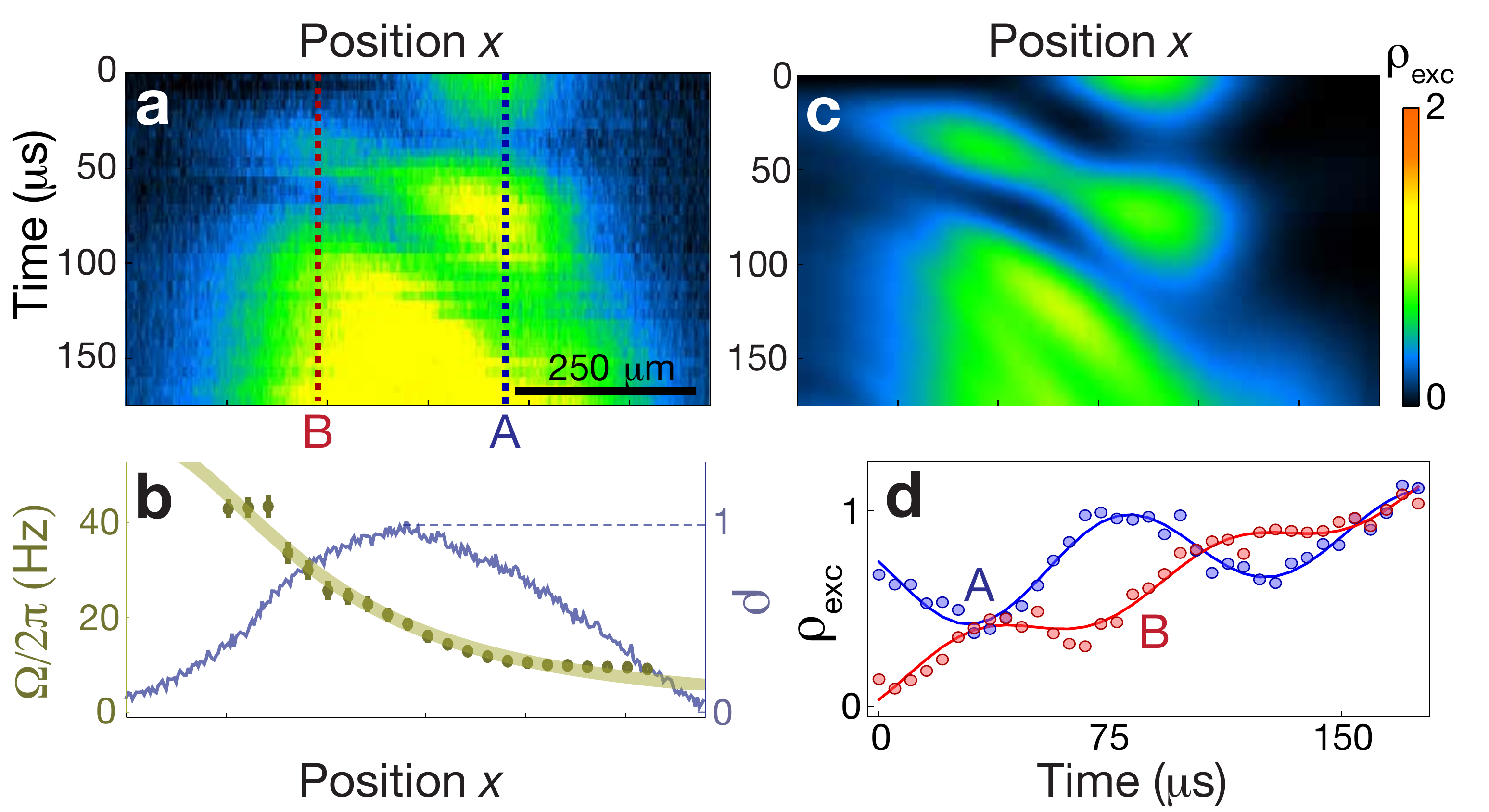}
\caption{\textbf{Cavity-mediated spin-exchange interactions.} (a) Driving the cavity induces spin excitations to hop from the right side of the atomic cloud (A) to the left (e.g., B) and back.  The vector light shift $\Omega(x)$ (b, yellow) and atomic density profile $\rho(x)$ (b, blue) serve as input to a mean-field model (c) of the spin dynamics. (d) Oscillations in excitation density $\rho_\mathrm{exc}$ vs time along cuts A and B; lines are guides to the eye.}\label{fig:xflipflop}
\end{figure}

To observe the photon-mediated interactions (Fig. \ref{fig:xflipflop}), we first initialize all atoms in $\ket{m_\f = -1}$ in a 4~G magnetic field.  We apply a local Raman $\pi/2$ pulse to populate a region of the cloud (A) with spin excitations.  At time $t=0$, we switch on the cavity drive field at a detuning $\delta_- = 2\pi\times 1.7$~MHz.  We observe the subsequent evolution of the spins by state-sensitive imaging \cite{SM}.  We regard the system as one-dimensional, integrating over the transverse dimensions of the atomic cloud and plotting the local density of spin excitations $\rho_{\text{exc}} = \rho(1 + \avg{f^z})$ vs time, where $\avg{\vec{f}}$ is the local spin polarization and $\rho$ is the local atomic density, normalized to peak density. The data show a coherent oscillation of spin excitations from the initially populated region (A) to elsewhere in the cloud (B) and back (Fig. \ref{fig:xflipflop}a, d).

A striking feature of the spin dynamics is their highly non-local character. The spin excitations first hop towards the left edge of the cloud, rather than to regions closer to the initially excited area (Fig. \ref{fig:xflipflop}a).  More generally, we observe a spatial gradient in the time-scale of the spin dynamics, which we attribute to a gradient in atom-light coupling: the coupling is strongest at the left because the atoms are displaced from cavity center.

To verify our understanding of the atom-light interactions, we have directly measured the ac Stark shift induced by the intracavity light as a function of position $x$ along the cavity axis.  Figure \ref{fig:xflipflop}b shows the on-axis vector light shift $\Omega(x) = [g_{m=-1}^2(x) - g_{m=1}^2(x)]/(2\Delta)$
per circularly polarized intracavity photon. The light shifts $\Omega_i\equiv \Omega(x_i)$ determine the spin-exchange couplings $\chi_{ij}^{\pm} = \bar{n}\Omega_i \Omega_j \disp(\delta_{\pm})/\kappa$, where $\bar{n}$ is the average intracavity photon number and $\disp(\delta) = \delta \kappa/(16[\delta^2 + (\kappa/2)^2])$ \cite{SM}. 



We use the measured light shift as input to a mean-field model (Fig. \ref{fig:xflipflop}c) with which we compare the observed spin dynamics. By reproducing the spatial structure of oscillations in the magnetization, the model corroborates the graph of nonlocal interactions $\chi_{ij}^{\pm}$.  The model also captures two dissipation mechanisms observed in the experiment: cavity decay induces spin relaxation towards $m_F = 1$, while inhomogeneous broadening due to the 5~$\micro{m}$ rms transverse cloud size causes additional damping \cite{SM}.





The effects of cavity decay visible in Figure \ref{fig:xflipflop}a can theoretically be reduced by increasing the detuning $\delta_\pm$ from Raman resonance. An optimal detuning $\delta_\mathrm{opt} \sim \sqrt{N\eta}\kappa$ is dictated by the collective cooperativity $N\eta = 4Ng^2/(\kappa\Gamma) \sim 10^6$, which quantifies two competing decay channels: collective decay via the cavity at small $\delta_\pm$ and spontaneous emission at larger $\delta_\pm$, where a stronger control field is required to maintain a fixed interaction strength.  Finite laser power currently limits us to small detunings $\delta_\pm \ll \delta_\mathrm{opt}$, leaving room to improve the interaction-to-decay ratio by a factor of $10^2$ in future experiments \cite{SM}.


Both the overall strength of the interactions and their sign are controlled by the drive laser.  However, the hopping dynamics of Fig. \ref{fig:xflipflop} do not reveal the sign of the couplings $\chi_{ij}$.  To more fully characterize the interactions, we note that the spin-exchange Hamiltonian can equivalently be rewritten as
\begin{equation}\label{eq:Hxy}
H = \sum_{i,j} \chi_{ij} \left(f_i^x f_j^x + f_i^y f_j^y \right) + \sum_i h_i f_i^z, 
\end{equation}
where we have set $\hbar=1$.  Here $\chi_{ij} = \chi_{ij}^+ + \chi_{ij}^-$, and $h_i = \chi_{ii}^+ - \chi_{ii}^-$.
By Eq. \ref{eq:Hxy}, each spin precesses about an effective field in the $xy$-plane generated by all other spins.  The rate and direction of the spin precession then reveal the magnitude and sign of $\chi_{ij}$.

\begin{figure}[htb]
\includegraphics[width=\columnwidth]{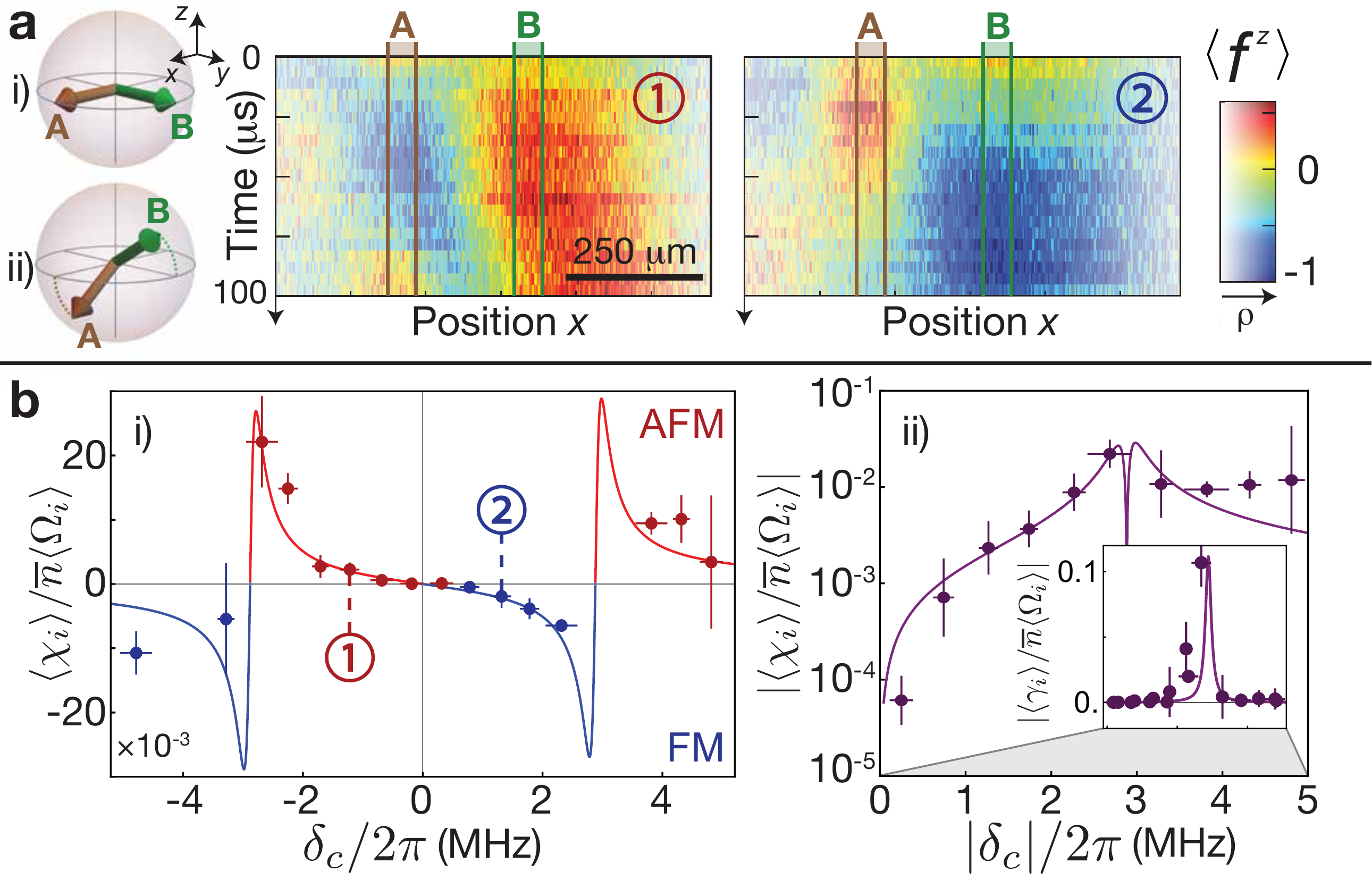}
\caption{\textbf{Optical control of spin-exchange interactions}. (a) Spins in regions A and B are initially oriented along $\hat{f}_x$ and $\hat{f}_y$, respectively.  Light-induced interactions convert this transverse polarization into a signal in $\langle f^z\rangle$, shown for two different drive frequencies.  Color scale indicates $\avg{f^z}$ (hue) and density $\rho$ (saturation).  (b) Varying the drive frequency changes the sign of interactions from antiferromagnetic (red) to ferromagnetic (blue). Solid curve is a fit with amplitude as the only free parameter. Right plot (ii) shows agreement of interaction strength $\abs{\avg{\chi_i}}$ with theory across two orders of magnitude.  Inset shows spin relaxation rate $\avg{\gamma_i}$.} \label{fig:strength_and_sign}
\end{figure}

To measure the couplings $\chi_{ij}$, we first prepare the non-interacting system with a spin texture in the $\hat{f}_x,\hat{f}_y$-plane (Fig. \ref{fig:strength_and_sign}a.i).  Using a pair of Raman pulses, we initialize one portion of the cloud (centered about region A in Fig. \ref{fig:strength_and_sign}a) with spins polarized along $\hat{f}_x$ and the remainder of the cloud (centered about region B) with spins polarized along $\hat{f}_y$, where the axes $\hat{f}_{x,y}$ are defined in a rotating frame at the Larmor frequency. By Eq. \ref{eq:Hxy}, cavity-mediated interactions should induce the $\hat{f}_x$- and $\hat{f}_y$-polarized spins to precess about one another in a direction that depends on the sign of $\chi_{ij}$ (Fig. \ref{fig:strength_and_sign}a.ii).  This precession converts the transverse spin texture into a signal in the longitudinal polarization $\langle f^z \rangle$.

The magnetization dynamics allow us to measure both the strength and sign of the flip-flop coupling as a function of drive frequency.  For ideal unitary dynamics, the initial rate of change $df_i^z/dt$ of each atom's magnetization would reveal its total coupling $\chi_i = \sum_{j}\chi_{ij}$ to all other spins. By comparing the initial slopes $\avg{d\f^z/dt}_{A,B}$ in regions A and B, and accounting for the calibrated spatial dependence of the atom-light coupling $\Omega(x)$, we extract both the mean spin relaxation rate $\langle \gamma_i \rangle$ and mean total coupling $\avg{\chi_i}$ within each region.





Figure \ref{fig:strength_and_sign}b compares the measured flip-flop coupling with theory.  Consistent with our expectation, the sign of the interaction changes as the drive frequency crosses through each of the Raman resonances $\ddelta = \pm \omega_Z$, and at the cavity resonance $\ddelta = 0$, where $\chi_{ij}^+ = - \chi_{ij}^-$.  This change in sign is evident in a striking reversal of the slopes of the magnetization vs time in regions A and B of the cloud.  The interaction strength per intracavity photon agrees with the independently calibrated atom-cavity coupling and also follows the predicted dependence on detuning over a wide dynamic range (Fig. \ref{fig:strength_and_sign}b.ii).  Lastly, the magnetization data confirm that the dissipation $\avg{\gamma_i}$ is highest on two-photon resonance ($\ddelta=\pm \omega_Z$).  





Whereas the spin-exchange dynamics considered above can be understood by regarding the spins as precessing about a classical mean field $\langle \vec{\f} \rangle$, the quantum system can exhibit dynamics even with zero average magnetization $\avg{\vec{\f}}=0$. To access dynamics driven by quantum fluctuations, we initialize an ensemble of atoms in $\ket{m_\f=0}$, which the flip-flop interactions can convert into correlated pairs of atoms in $\ket{m_\f =\pm 1}$ (Fig. \ref{fig:exp_setup}b.iii). This spin mixing process is analogous to an optical parametric oscillator, with the large population $N_0$ of $\ket{m_\f=0}$ atoms serving as a pump.


The spin mixing can thus be understood by viewing the atomic populations in $m_\f=\pm1,0$ as excitations of three bosonic modes $a,b,c$. In the limit of uniform coupling $\chi_{ij} \sim \chi$, we can rewrite the spin operators in Eq. (\ref{eq:Hflipflop}) in terms of these modes:
\begin{equation}
H_{\mathrm{mix}} = 2 \chi c^2 a^\dagger b^\dagger + \mathrm{h.c.} + H_q,
\label{eq:threemodemodel}
\end{equation}
where the first term is responsible for pair creation and $H_q = (2 \chi c^\dagger c+q+\chi) (a^\dagger a + b^\dagger b + 1)$ includes a quadratic Zeeman shift $q/B^2 = 2\pi \times 144$~Hz/$\mathrm{G}^2$ that can suppress pair creation. Instability to the production of pairs occurs when the collective interaction strength $4N_0 \chi$ is larger in magnitude than the quadratic Zeeman shift and has opposite sign \cite{SM}, as observed in ferromagnetic spinor condensates \cite{sadler2006spontaneous,klempt2010parametric,linnemann2016quantum,stamper2013spinor,luo2017deterministic}.

\begin{figure}[htb]
\includegraphics[width=\columnwidth]{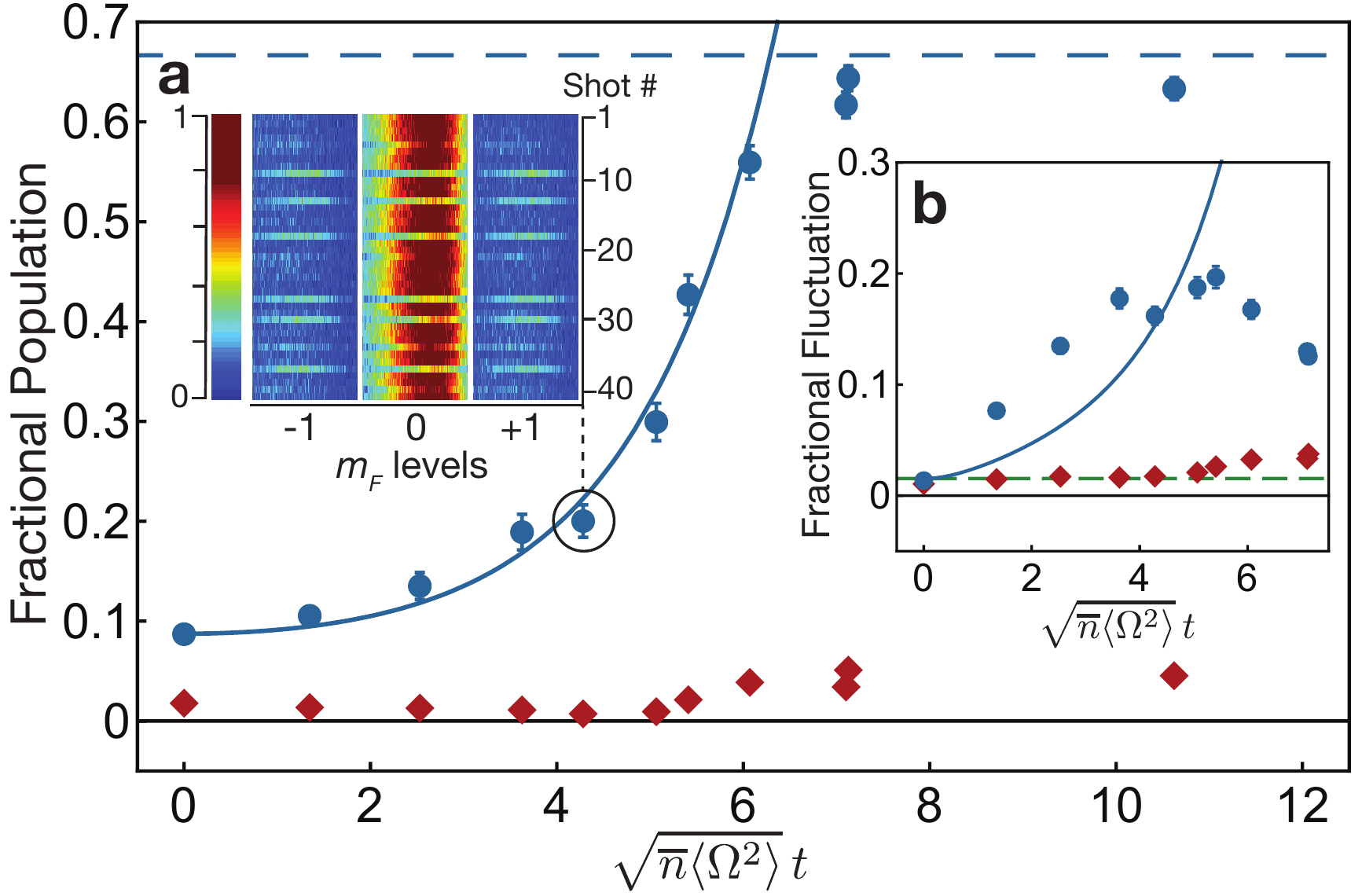}
\caption{\textbf{Cavity-mediated spin mixing} in a cloud of $N = 10^5$ atoms. (a) Average side mode population $N_s$ (blue circles) and population difference $F_z$ (red diamonds) vs $\sqrt{\overline{n}\avg{\Omega^2}} t$ \cite{SM}, measured for interaction times $0 \leq t\leq 1.2$~ms with typical intracavity photon number $\overline{n}\approx 3\times 10^3$. 
Inset: images from 40 iterations of the experiment with $t = 400~\mu s$; colors indicate fractional population in each state. (b) Fluctuations in side mode population $\Delta N_s$ (blue circles) and population difference $\Delta F_z$ (red diamonds). Solid blue curves are obtained by fitting $N_s$ with the model in Eq. (\ref{eq:side}) and plotting the corresponding prediction for $\Delta N_s$ with no free parameters. Dashed blue line in (a) indicates saturation level $N_s/N = 2/3$ for the side mode population. Dashed green line in (b) indicates detection noise.}\label{fig:spin_mixing}
\end{figure}

To enable cavity-mediated pair creation, we initialize nearly all atoms in $\ket{m_\f=0}$ in a weak magnetic field $B=1.14$~G and induce ferromagnetic interactions with a red-detuned drive field.  After driving the cavity at Raman detuning $\delta_- = -2\pi \times 600$~kHz for a variable time $t\leq 1.2~\mathrm{ms}$, we image the populations of the three Zeeman states.  Figure \ref{fig:spin_mixing}a shows representative images from 40 iterations of such an experiment, with $t = 400~\micro{s}$.  We observe a macroscopic population of the $m_\f = \pm 1$ ``side modes'' (Fig. \ref{fig:spin_mixing}), with large shot-to-shot fluctuations that are well correlated between the $m_\f=\pm 1$ states.

The rapid growth in total side mode population $N_s$ at fixed population difference $F_z = a^\dagger a - b^\dagger b$ is qualitatively consistent with the parametric amplification model $H_{\mathrm{mix}}$.  In the experimentally relevant limit $\abs{\chi} \ll q \ll N_0 \abs{\chi}$, this model predicts an initial population growth
\begin{equation}\label{eq:side}
N_s(t) =\left[ \frac{4 N_0 \chi}{\lambda} \right]^2 \left( N_s(0) + 1 \right) \left( \cosh \lambda t - 1 \right) + N_s(0),
\end{equation}
where $\lambda = 4 \sqrt{N_0 q |\chi|}$ and $N_s(0)$ represents initial population in the side modes, present in the experiment due to imperfect state preparation.  Fitting Eq. (\ref{eq:side}) to the early-time population dynamics in the experiment (Fig. \ref{fig:spin_mixing}, solid blue) yields a time constant $1/\lambda = 160(20)~\micro{s}$ for the exponential growth, which is six times slower than expected for a system with uniform coupling equal to the rms coupling $\sqrt{\langle\Omega^2\rangle}$ in our system.  The slower growth we observe may be due to additional effects of inhomogeneity or residual population in hyperfine states not included in the three-mode model.




The parametric amplification model predicts macroscopic fluctuations in side mode population, with $\Delta N_s \approx N_s/\sqrt{2}$ at early times.  While the overall scale of the fluctuations that we observe roughly matches this expectation (Fig. \ref{fig:spin_mixing}b), a more detailed analysis remains a subject for future investigation.  Of particular interest are the fluctuations in population difference $F_z$, which for ideal unitary pair creation should remain zero.  The measured fluctuations in $F_z$ at short times are currently dominated by percent-level technical noise in state preparation and detection (Fig. 4b, dashed green line).  Reducing this technical noise---or harnessing interaction-based readout \cite{davis2016approaching,hosten2016quantum,linnemann2016quantum}---will allow for probing entanglement between the $m_\f = \pm 1$ modes \cite{lucke2014detecting}, enabling applications in quantum metrology \cite{masson2017cavity}.

Notably, light-mediated spin mixing will allow for generating spin nematic squeezing and twin Fock states significantly faster than in past experiments harnessing contact interactions \cite{masson2017cavity,chang2005coherent,sadler2006spontaneous,klempt2010parametric,lucke2011twin,hamley2012spin,luo2017deterministic}. The optically controlled interactions further allow for probing ferromagnetic and antiferromagnetic spinor phases in a single atomic species, and for tuning the interaction range \cite{hung2016quantum}.



The combination of non-local spin interactions with local addressing and imaging opens the door to controlling and probing the spatial structure of entanglement.  Applications range from quantum-enhanced magnetic field imaging to investigating fundamental limits on the propagation of quantum correlations \cite{bentsen2018fast}.  Quantum optical approaches to combinatorial optimization problems \cite{torggler2018quantum,gopalakrishnan2011frustration}, e.g., number partitioning \cite{mertens1998phase, lucas2014ising}, could be explored by positioning individual spins to specify their interactions.  Extensions of our scheme will allow for engineering a wider range of non-local graphs \cite{hung2016quantum}, enabling exotic long-range interactions that can stabilize topological order \cite{hung2016quantum} or mimic toy models of quantum gravity \cite{gubser2017p,*gubser2018continuum}.  

\begin{acknowledgments}
This work was supported by the National Science Foundation under grant No. PHY-1506401 and by the Research Corporation Cottrell Scholars Program.  E. J. D. and G. B. acknowledge support from the National Science Foundation Graduate Research Fellowship Program.  E. J. D. acknowledges support from the Hertz Foundation.  L. H. acknowledges support from the Max Weber Foundation.  T. Li was supported by the Air Force Office of Scientific Research.  We thank R. Yanagimoto for technical assistance.
\end{acknowledgments}

\bibliography{flipflop}

\clearpage


\section*{Supplementary material}

# Photon-Mediated Spin-Exchange Dynamics of Spin-1 Atoms: Supplemental Material

(Dated: December 15, 2018)

In this supplement, we elaborate on derivations of theoretical models in the main paper and on details from the experimental sequence. In Sec. I, we derive the spin-exchange Hamiltonian, elaborate on the mean-field model of the spin dynamics, and describe how to relate the strength of the interaction $\chi$ to parameters measured in the experiment. Additionally, we derive the short-time behavior for the spin-mixing Hamiltonian. In Sec. II, we elaborate on the experimental methods and describe the calibration of the vector light shift $\Omega(x)$ by Ramsey spectroscopy.

## I. THEORETICAL BACKGROUND

### A. Derivation of Spin-Exchange Hamiltonian

We here derive the spin-exchange Hamiltonian arising in a driven optical cavity with a magnetic field **B** perpendicular to the cavity axis. We consider atoms with hyperfine spin **f** and let **B** define the quantization axis for the spins, so that $f_i^z$ denotes the spin projection of the $i^{\text{th}}$ atom along the magnetic field. The spins couple to the cavity mode via the Faraday effect [1, 2], which is most conveniently described by defining the Stokes vector **S** of the intracavity light. For the light, we choose the *cavity axis* as the quantization axis, defining

$$S_z = \left(a_+^\dagger a_+ - a_-^\dagger a_-\right)/2, \tag{S1}$$

where $a_\pm$ are the annihilation operators for $\sigma_\pm$-polarized cavity modes. The Faraday interaction then takes the form

$$H_I = \Omega S_z \mathcal{F}_x, \tag{S2}$$

where $\mathcal{F} = \sum_i \xi_i \mathbf{f}_i$ denotes the collective spin of the ensemble, weighted according to factors $\Omega \xi_i = g_i^2/\Delta_i$ that account for any inhomogeneity in the vacuum Rabi frequency $g_i$, or of the atomic transition frequency $\omega_i = \omega_c^N - \Delta_i$, where $\omega_c^N$ denotes the cavity mode frequency in the presence of $N$ atoms. The Faraday interaction generates a rotation in the cavity mode polarization by an angle proportional to the 'magnetic field' produced by the atoms along $\mathcal{F}_x$.

The microscopic mechanism for the flip-flop dynamics is more evident if we rewrite the Faraday interaction in terms of operators $\mathcal{H}, \mathcal{V}$ representing the two orthogonal linearly polarized cavity modes:

$$\mathcal{H} = \frac{a_+ + a_-}{\sqrt{2}} \tag{S3}$$

$$\mathcal{V} = \frac{a_+ - a_-}{i\sqrt{2}}. \tag{S4}$$

In terms of the linearly polarized modes, the Stokes vector has $z$-component

$$S_z = \frac{\mathcal{V}^\dagger \mathcal{H} - \mathcal{H}^\dagger \mathcal{V}}{2i}, \tag{S5}$$

yielding an interaction Hamiltonian

$$H_I = \frac{\Omega}{4i} \left(\mathcal{V}^\dagger \mathcal{H} - \mathcal{H}^\dagger \mathcal{V}\right)\left(\mathcal{F}_+ + \mathcal{F}_-\right). \tag{S6}$$

Thus, $H_I$ describes processes wherein an atom flips its spin and transfers a photon between the two linearly polarized cavity modes. These Raman processes are resonant when the drive field is tuned to a frequency $\omega_c^N \pm \omega_Z$. More generally, tuning the drive field to a frequency $\omega_d = \omega_c^N + \delta_c$ results in detunings $\delta_\pm \equiv \delta_c \mp \omega_Z$ from the two Raman resonances.

The dynamics in the driven cavity are described by a master equation with a Hamiltonian of the form

$$H = H_I + \omega_Z F_z + \sum_\pm \left[\omega_c a_\pm^\dagger a_\pm + \varepsilon_\pm e^{-i\omega_d t} a_\pm^\dagger + \varepsilon_\pm^* e^{i\omega_d t} a_\pm\right], \tag{S7}$$

where $\mathbf{F} = \sum_i \mathbf{f}_i$ denotes the uniformly weighted collective spin that couples to the magnetic field and $\varepsilon_\pm$ describe coherent fields driving the cavity on the $\sigma_\pm$ polarized cavity modes. We will analyze the time evolution in an interaction picture wherein states evolve with the trivial portion of the Hamiltonian,

$$H_0 = \omega_Z F_z + \sum_\pm \left[ \omega_c a_\pm^\dagger a_\pm \right]. \tag{S8}$$

The remaining dynamics, encoded in the operators, are governed by an effective Hamiltonian $\tilde{H} = UHU^\dagger - H_0$, where $U = e^{-iH_0 t}$. This transformation yields

$$\tilde{H} = \tilde{H}_I + \tilde{H}_{\text{drive}}, \tag{S9}$$

where

$$\tilde{H}_I = \frac{\Omega}{4i} \left( \mathcal{V}^\dagger \mathcal{H} - \mathcal{H}^\dagger \mathcal{V} \right) \left( \mathcal{F}_+ e^{i\omega_Z t} + \mathcal{F}_- e^{-i\omega_Z t} \right), \tag{S10}$$

and

$$\tilde{H}_{\text{drive}} = \sum_\pm \varepsilon_\pm e^{-i\delta_c t} a_\pm^\dagger + \text{h.c.}. \tag{S11}$$

We will henceforth remain in the interaction picture and drop the tilde for notational simplicity.

If we include cavity decay in the model, the time evolution of each operator $\mathcal{O}$ is described by a master equation in Lindblad form,

$$\dot{\mathcal{O}} = i[H, \mathcal{O}] + \frac{\kappa}{2} \sum_\pm \left( [a_\pm^\dagger, \mathcal{O}] a_\pm + a_\pm^\dagger [\mathcal{O}, a_\pm] \right). \tag{S12}$$

In the absence of atoms, the cavity field operators thus evolve as

$$\dot{a}_\pm = -\frac{\kappa}{2} a_\pm - i\varepsilon_\pm e^{-i\delta_c t}. \tag{S13}$$

The steady-state field is given by

$$\bar{a}_\pm(t) = \frac{-i\varepsilon_\pm e^{-i\delta_c t}}{\kappa/2 - i\delta_c}. \tag{S14}$$

We will focus on the case of a cavity mode driven with horizontally polarized light, such that the steady-state fields in the linear basis (Eqs. S3-S4) are $\overline{\mathcal{V}} = 0$ and

$$\overline{\mathcal{H}} = \frac{-i\varepsilon e^{-i\delta_c t}}{\kappa/2 - i\delta_c}, \tag{S15}$$

where $\varepsilon = (\varepsilon_+ + \varepsilon_-)/\sqrt{2}$. Then the average number of photons in the driven cavity mode is

$$\bar{n} \equiv |\overline{\mathcal{H}}|^2 = \frac{|\varepsilon|^2}{\delta_c^2 + (\kappa/2)^2}. \tag{S16}$$

It will be useful to write the field operators in terms of the $c$-numbers $\overline{\mathcal{H}}$ and $\overline{\mathcal{V}}$ and the fluctuations $\hat{h}, \hat{v}$ about these values:

$$\hat{\mathcal{H}} = \overline{\mathcal{H}} + \hat{h}, \tag{S17}$$
$$\hat{\mathcal{V}} = \overline{\mathcal{V}} + \hat{v}, \tag{S18}$$

where we have temporarily written explicit hats to emphasize which symbols are operators.

For small fluctuations about the steady-state field values, we can solve for the spin dynamics by including only terms to lowest order in $\hat{h}$ and $\hat{v}$ in the interaction Hamiltonian. The atom-light interaction can thus be approximated as

$$H_I \approx \frac{\Omega/4}{\sqrt{\delta_c^2 + (\kappa/2)^2}} \left( \varepsilon e^{-i\delta_c t - i\phi} v^\dagger + \varepsilon^* e^{i\delta_c t + i\phi} v \right) \left( \mathcal{F}_+ e^{i\omega_Z t} + \mathcal{F}_- e^{-i\omega_Z t} \right). \tag{S19}$$

where $\phi$ is an unimportant phase factor that accounts for the phase delay between the drive field and the cavity field. Here, we can see that resonant spin-flip processes occur only for a laser detuned from cavity resonance by $\delta_c = \pm\omega_Z$. However, for arbitrary drive frequency $\delta_c = \pm\omega_Z + \delta_\pm$, resonant *pairwise* flip-flops can arise from $H_I$ in second-order perturbation theory. To further analyze the spin dynamics, we will adiabatically eliminate the $\hat{v}$ cavity mode by assuming a weak drive $\varepsilon$ and large detuning $\delta_\pm$ from Raman resonance.

By adiabatically eliminating the cavity modes [3], we arrive at an effective Hamiltonian

$$H_{\text{eff}} = \frac{|\Omega\varepsilon/4|^2}{\delta_c^2 + (\kappa/2)^2} \sum_\pm \left( \frac{\delta_\pm}{\delta_\pm^2 + (\kappa/2)^2} \mathcal{F}_\pm \mathcal{F}_\mp \right). \tag{S20}$$

The effective Hamiltonian can be written in the form

$$H_{\text{eff}} = \sum_{i,j} \chi_{ij} \left( f_i^x f_j^x + f_i^y f_j^y \right) + \sum_i h_i f_i^z \tag{S21a}$$

with the coupling constants $\chi_{ij} = \chi_{ij}^+ + \chi_{ij}^-$ given by

$$\chi_{ij}^\pm = \bar{n} \frac{\Omega_i \Omega_j}{16} \frac{\delta_\pm}{\delta_\pm^2 + (\kappa/2)^2}. \tag{S21b}$$

Here, $\Omega_i = \Omega\xi_i$ represents the vector light shift imparted by a circularly polarized intracavity photon to the $i^{\text{th}}$ atom. The extra 'magnetic field' term is generated by commutators of $f_i^x, f_i^y$, and has the form:

$$h_i = \bar{n} \frac{\Omega_i^2}{16} \sum_\pm \frac{(\pm 1)\delta_\pm}{\delta_\pm^2 + (\kappa/2)^2}. \tag{S21c}$$

In the case of uniform coupling, we can ignore this longitudinal field in the dynamics by going into a rotating frame. More generally, this term contributes some dephasing between spins at different positions in the cloud.

The Hamiltonian dynamics are generically accompanied by dissipation described by an effective Lindblad operator

$$L_v = \frac{\Omega\varepsilon}{4} \frac{\sqrt{\kappa}}{\delta_c - i\kappa/2} \sum_\pm \frac{\mathcal{F}_\pm e^{-i\delta_\pm t - i\phi}}{\delta_\pm + i\kappa/2}. \tag{S22}$$

The Lindblad operator consists of two terms that rotate rapidly with respect to one another. We can therefore break these two terms up into separate Lindblad operators:

$$L_\pm = \sqrt{\gamma_\pm} \mathcal{F}_\pm, \tag{S23a}$$

where

$$\gamma_\pm = \bar{n} \frac{\Omega^2}{16} \frac{\kappa}{\delta_\pm^2 + (\kappa/2)^2} \tag{S23b}$$

and we have eliminated phase factors that are irrelevant to the dynamics. Comparing equations S21b and S23b, we see that the interaction-to-decay ratio

$$\frac{\chi_{ij}^\pm}{\gamma_\pm} = \frac{\delta_\pm}{\kappa} \xi_i \xi_j \tag{S24}$$

improves with increasing detuning $\delta_\pm$ from Raman resonance.

### B. Dynamics Including the Light Field

Even in the regime where the adiabatic elimination is not valid, we can write down a full set of Heisenberg equations of motion including the cavity field operators as well as the spin operators. In particular, from the master equation

(Eq. S12) for the Hamiltonian in Eq. S7 with $\mathcal{H}$-polarized drive field, the Heisenberg equations are

$$\dot{\mathcal{V}} = -\Omega \mathcal{F}^x \mathcal{H}/2 - \left(\frac{\kappa}{2} - i\delta_c\right)\mathcal{V} \tag{S25a}$$

$$\dot{\mathcal{H}} = \Omega \mathcal{F}^x \mathcal{V}/2 - \left(\frac{\kappa}{2} - i\delta_c\right)\mathcal{H} - \varepsilon \tag{S25b}$$

$$\dot{f}_i^x = -\omega_Z f_i^y \tag{S25c}$$

$$\dot{f}_i^y = \omega_Z f_i^x - \Omega_i S_z f_i^z \tag{S25d}$$

$$\dot{f}_i^z = \Omega_i S_z f_i^y. \tag{S25e}$$

Here $S_z$ is the component of the Stokes vector along the cavity axis, which is given in terms of $\mathcal{H}$ and $\mathcal{V}$ in Eq. S5.

While Equations S25 are exact, evaluating the exact time evolution including atom-light correlations is non-trivial. As a lowest-order approximation, we employ a mean-field treatment, calculating the approximate time-evolution of the expectation values $\langle\mathcal{V}\rangle, \langle\mathcal{H}\rangle$, and $\langle\mathbf{f}_i\rangle$ by neglecting atom-field correlations on the right-hand side of Eqs. S25.

We numerically solve Eqs. S25 in the mean-field approximation to obtain Fig. 2c in the main text, using the measurement of the vector light shift $\Omega(x)$ and the initial atomic state as inputs. The drive-cavity detuning $\delta_c = -2\pi \times 0.875$ MHz is fit by eye to match the data qualitatively and agrees within error with the value set during the experiment, $\delta_c = -2\pi \times 1.1(3)$ MHz. The drive strength $\varepsilon = 2\pi \times 95$ MHz used in the simulation corresponds to $\bar{n} = 12 \times 10^3$ intracavity photons (Eq. S16), consistent with the value $\bar{n} = 9(2) \times 10^3$ inferred from the cavity transmission.

Our simulation accounts for two different dissipation mechanisms that affect the coherence of the spin-exchange oscillations, namely, thermal broadening and cavity decay. These contributions are examined in detail in Fig. S1, where part (a) shows the model and data from Fig. 2 of the main text in terms of local magnetization $F_z \equiv \rho\langle f^z\rangle$, where $\langle f^z\rangle$ denotes local spin polarization and $\rho$ denotes local density. At fixed longitudinal position $x_i$, thermal

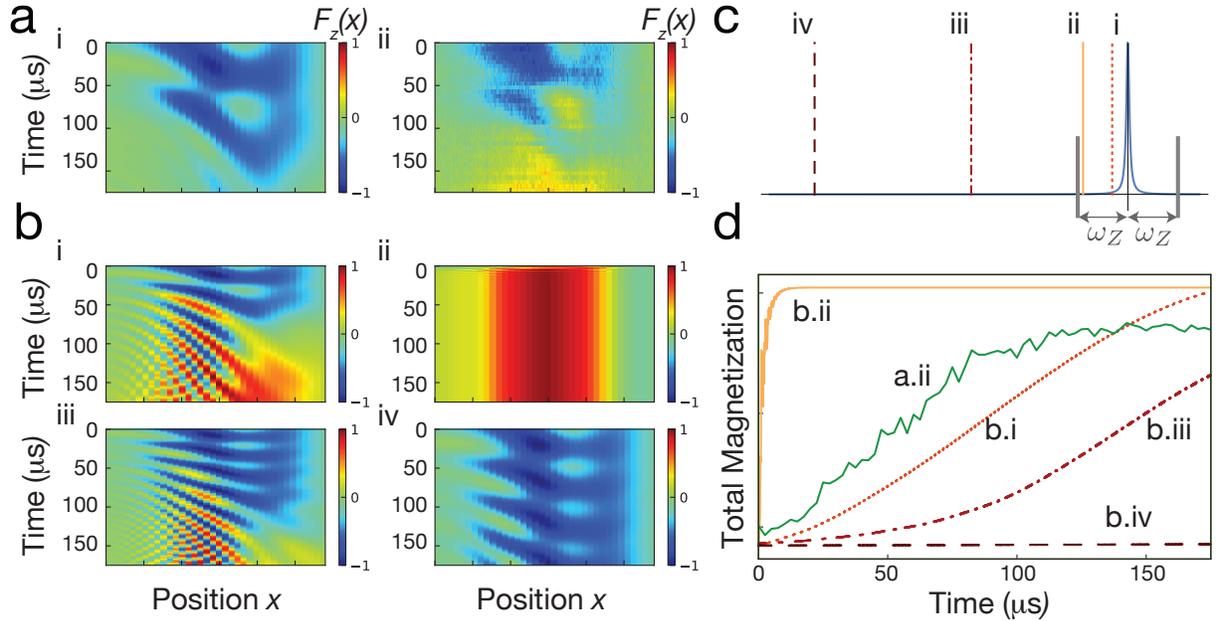

FIG. S1: **Effect of thermal broadening and supperradiant decay on the unitary dynamics.** (a) The model including thermal broadening (i) washes out the oscillations as in the experimental data (ii) compared to the zero-temperature model at the same detuning $\delta_- = 2\pi \times 1.93$ MHz (b.i), but artificially suppresses the relaxation rate compared to the data and the zero-temperature model (b.i). (b) The zero-temperature model shows rapid relaxation near the Raman resonance $\delta_- = 2\pi \times .30$ (ii) and slower relaxation at larger detunings $\delta_- = -2\pi \times 5.95, -2\pi \times 14.7$ MHz (iii, iv). (c) Schematic of drive detunings with respect to cavity and Raman resonances. (d) The total magnetization $\sum_x F_z(x)$ over the cloud is plotted vs time. Green curve corresponds to the data in (a)ii; solid orange-dashed burgundy curves correspond to calculated dynamics in (b)i-iv at the drive frequencies indicated by the lines in (c).

broadening results in a range of radial couplings $\Omega_i(r)$. The distribution of radial couplings is averaged over in our data (Fig. S1a.ii). To incorporate the effect of thermal broadening in the simulated dynamics, we average the dynamics over a Boltzmann distribution of atoms in the trap, assuming a Gaussian cloud of atoms in a three-dimensional harmonic well at each lattice site. The oscillations are washed out in a thermal cloud (Fig. S1a.i) compared to the dynamics at zero temperature (S1b.i) with all other parameters held fixed. The second mechanism is superradiant decay due to loss of cavity photons, which manifests as a spin relaxation rate towards either $m_f = \pm 1$, depending on the Raman detunings $\delta_\pm$. In Fig. S1b.i-iv, we plot the zero-temperature dynamics at detunings $\delta_- = 2\pi \times 1.93, 2\pi \times .30, -2\pi \times 5.95, -2\pi \times 14.7$ MHz (Fig. S1c) at fixed intracavity photon number. The relaxation rate is highest near the Raman resonance $\delta_- = 0$, but is suppressed for larger detunings, as summarized in Fig. S1d showing the total magnetization $\sum_x F_z(x)$ vs time.

Compared with the zero-temperature models and with our experimental data, the model including thermal broadening (Fig. S1a.i) shows a reduced relaxation rate. We attribute this to the fact that the static inhomogeneous couplings assumed in the model can lead to the formation of a dark state, which would decohere in the actual experiment due to atomic motion. Nevertheless, the key features of our data are well captured by the combination of these two models.

### C. Spin Mixing Dynamics

Here we derive the short-time equations of motion for the spin-mixing dynamics in the limit where atoms are uniformly coupled to the cavity mode, applying a mean-field treatment. We show that the results are equivalent to the well known mean-field treatments of spin-mixing in spinor Bose-Einstein condensates [4].

To model the spin mixing dynamics, we use a three-mode Schwinger boson representation for the collective spin $\mathbf{F}$ of the atomic ensemble:

$$F^+ = \sqrt{2}\left(a^\dagger c + c^\dagger b\right) = \left(F^-\right)^\dagger \tag{S26a}$$
$$F_z = a^\dagger a - b^\dagger b. \tag{S26b}$$

We obtain a Hamiltonian similar in form to a degenerate parametric oscillator with pump mode $c$ ($m_f = 0$) and side modes $a, b$ ($m_f = \pm 1$):

$$H_{\rm eff} = 2\chi\left[c^\dagger c^\dagger ab + a^\dagger b^\dagger cc + a^\dagger a(1 + c^\dagger c) + c^\dagger c(1 + b^\dagger b)\right] + ha^\dagger a - hb^\dagger b + q(a^\dagger a + b^\dagger b), \tag{S27}$$

where $\chi = \chi^+ + \chi^-$, $h = \chi^+ - \chi^-$, and $q$ represents the quadratic Zeeman shift. Then, by treating the pump mode classically, $\hat{c} \to c e^{i\phi}$ and $\hat{c}^\dagger \to c e^{-i\phi}$ we obtain a quadratic Hamiltonian in the side mode operators $a, b$:

$$H_{\rm eff} = 2\chi\left[c^2 e^{-2i\phi} ab + c^2 e^{2i\phi} a^\dagger b^\dagger + a^\dagger a(1 + c^2) + c^2(1 + b^\dagger b)\right] + ha^\dagger a - hb^\dagger b + q(a^\dagger a + b^\dagger b) \tag{S28}$$

We can conceptually simplify this further by introducing the bilinears

$$K_z = (a^\dagger a + b^\dagger b + 1)/2 \tag{S29a}$$
$$K^+ = a^\dagger b^\dagger = (K^-)^\dagger \tag{S29b}$$

where $2K_z - 1 = N_s$ is the total side-mode population and $K^\pm$ are 'pair-creation / annihilation' operators. The operators $\{K_z, K^\pm\}$ form a closed $SU(1,1)$ algebra [5–7]:

$$[K_z, K^\pm] = \pm K^\pm \tag{S30a}$$
$$[K^+, K^-] = -2K_z \tag{S30b}$$

The magnetization $F_z$ commutes with all of the operators $K^\pm, K_z$, so we can simply treat it as a constant $c$-number.

The group $SU(1,1)$ is very similar to the more familiar group $SU(2)$. In fact, the only difference is the minus sign in the second row of Eq. (S30b). While $SU(2)$ generates rotations in 3-dimensional Euclidean space, the group $SU(1,1)$ generates rotations and boosts in $(2+1)$-dimensional Minkowski space. In terms of $\mathbf{K}$, we can write the Hamiltonian as:

$$H_{\rm eff} = \mathbf{M} \cdot \mathbf{K} \tag{S31}$$

where $K_x = (K^+ + K^-)/2$, $K_y = (K^+ - K^-)/2i$, and

$$\mathbf{M} = \langle 4c^2\chi\cos 2\phi, -4c^2\chi\sin 2\phi, \chi(2 + 4c^2) + 2q\rangle \tag{S32}$$

and where we have dropped an overall constant term $(\chi + h)F_z - \chi - q$.

We solve for the motion of the system assuming that the pump field is static (i.e. $c, \phi$ are time-independent). By performing a rotation in the $x - y$ plane we can always pick a coordinate system for which $\phi = 0$. Then our equations of motion are:

$$\frac{d}{dt} \begin{bmatrix} K_x \\ K_y \\ K_z \end{bmatrix} = \begin{bmatrix} 0 & -M_z & 0 \\ M_z & 0 & M_x \\ 0 & M_x & 0 \end{bmatrix} \begin{bmatrix} K_x \\ K_y \\ K_z \end{bmatrix} \tag{S33}$$

whose eigenvalues are 0 and $\pm\lambda = \pm\sqrt{M_x^2 - M_z^2}$, with $M_x = 4c^2\chi$, $M_z = \chi(2 + 4c^2) + 2q$. For $|M_z| < |M_x|$, the eigenvalue $\lambda$ is real and positive, resulting in the dynamical instability that we observe in our experiment. In terms of the net quadratic Zeeman shift $\tilde{q} = q + \chi$ and the number of "pump" atoms $c^2 \sim N_0$ in the $m = 0$ mode, the condition for instability is,

$$\left| 1 + \frac{\tilde{q}}{2N_0\chi} \right| < 1 \tag{S34}$$

which is satisfied when $\tilde{q}$ and $\chi$ have opposite signs and the collective interaction strength $4N_0|\chi|$ exceeds the quadratic Zeeman shift. These are the same conditions for instability found for spin-mixing dynamics in spinor BECs.

To compute the time-dependent expectation value and standard deviation of $N_s(t)$, we assume that the system starts in a number state $|n_a, n_b, n_c\rangle$. We then obtain:

$$\langle N_s(t) \rangle = \alpha^2 \left(N_s(0) + 1\right)(\cosh \lambda t - 1) + N_s(0) \tag{S35}$$

and

$$\Delta N_s(t) = \sqrt{\langle N_s^2(t) \rangle - \langle N_s(t) \rangle^2} = (N_s(0) + 1)\sqrt{\alpha^2 \left[(\alpha^2 - 1)(\cosh \lambda t - 1)^2 + \sinh^2 \lambda t\right]/2} \tag{S36}$$

where

$$\lambda = 2\sqrt{-\tilde{q}(\tilde{q} + 4c^2\chi)}, \tag{S37}$$

$$\alpha = \frac{4c^2|\chi|}{\lambda}. \tag{S38}$$

Although these equations of motion are applicable for perfectly uniform couplings $\Omega(x)$, accounting to lowest order for the non-uniformity of the couplings shows that the growth rate scales as $\xi_{\rm rms} = \sqrt{\sum_i \xi_i^2}$. We account for this effect in the main text by plotting the growth in side-mode population in Fig. 4 as a function of $\sqrt{\langle \Omega^2 \rangle} = \Omega \xi_{\rm rms}$.

## II. EXPERIMENTAL METHODS

### A. Optical Cavity and Trapped Atoms

Our optical cavity has a near-concentric geometry with a maximum length $L_0 = 5$ cm. The precise length, as well as the mirror alignment, are fine-tuned with piezoelectric positioning stages. Our experiments were conducted at a cavity length $L = L_0 - 80$ $\mu$m, which results in a mode waist of $w_{780} = 16$ $\mu$m for the 780-nm drive field. The mirrors are additionally coated for 1560-nm light that is used to trap the atoms in one-dimensional lattice, with transverse confinement set by the waist $w_{1560} = \sqrt{2}w_{780}$. A typical trap depth during the experiment is $h \times 20$ MHz at the center of the cavity. The atomic cloud has an RMS width of 5 $\mu$m in the radial direction, is approximately 500 $\mu$m in length, and is centered about one Rayleigh length from cavity center.

### B. State Preparation and Detection

We provide here further details about the experimental sequence, including the state-sensitive imaging used to obtain measurements of the population in the three different Zeeman states. The atoms are loaded into the 1560-nm lattice from a 3D MOT, then placed in $|f = 1, m_f = -1\rangle$ by optical pumping into $|2, -2\rangle$ on the D1 line followed by an adiabatic microwave sweep. Additional microwave sweeps or optical Raman pulses are used to prepare the initial

Zeeman states for the subsequent quench dynamics. For the quench dynamics of Figs. 2 and 3 [or Fig. 4] in the main text, we apply a magnetic field $B_z = 4.0$ G [or $B_z = 1.0$ G], which adds to an ambient field $B_x = 0.52$ G, $B_y = -0.14$ G (axes as depicted in Fig. 1 of the main text). After the quench dynamics in the $f = 1$ hyperfine manifold, we detect the $|1,-1\rangle$ atoms by performing an adiabatic sweep on the $|1,-1\rangle \to |2,-2\rangle$ transition in a 4 G magnetic field and imaging on the cycling transition. We then detect the $|1,1\rangle$ atoms by performing a microwave sweep on the $|1,1\rangle \to |2,2\rangle$ transition and imaging on the cycling transition. Finally, we image the remaining atoms in $|1,0\rangle$ by adding repumping light.

### C. Measurement of Cavity Coupling

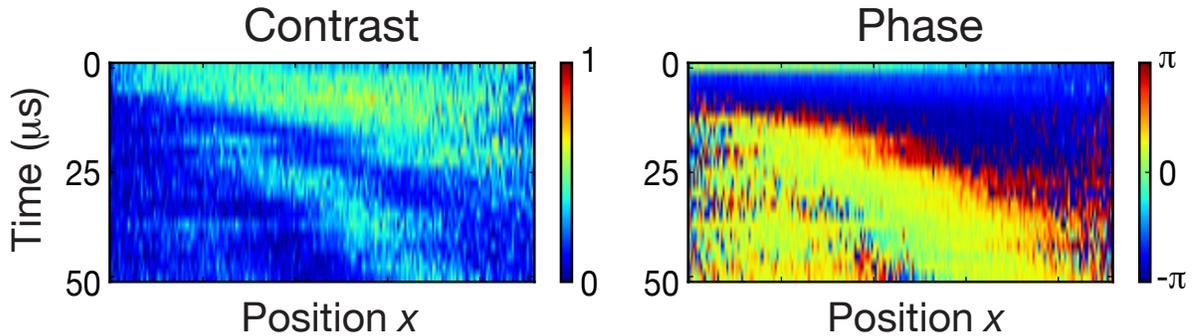

FIG. S2: Contrast and phase of the magnetization from the Ramsey measurement of the vector light shift. We measure the two quadratures $f_i^z(0), f_i^z(\pi/2)$ as a function of time, and can thus extract the local phase $\mathrm{Arg}(f_i^z(0) + if_i^z(\pi/2))$ and contrast $\sqrt{(f_i^z(0))^2 + (f_i^z(\pi/2))^2}$.

We measured the vector light shift $\Omega(x)$ (Fig. 2b in the main text) by performing Ramsey spectroscopy in the $f = 1$ hyperfine manifold. After initializing the atoms in $|1,-1\rangle$ with respect to a magnetic bias field along the cavity axis, we first apply a Raman $\pi/2$ pulse that rotates the Zeeman spin into the $f_x, f_y$-plane. We then drive the cavity with a variable-length pulse of circularly polarized light that modifies the Larmor precession rate, before performing a second $\pi/2$ pulse and reading out the local magnetization $f_i^z$. Fig. S2 shows the contrast and phase of the Ramsey fringe as a function of the duration of the light pulse (Fig. S2). In the ideal case where the light shift $\Omega$ depends only on the position $x_i$, the magnetization would simply evolve as $f_i^z(t)/f_i = e^{-i\bar{n}\Omega_i t}$. However, our data shows a decaying contrast due to inhomogeneous broadening caused by a thermal distribution of radial couplings $P(\Omega_i(r))$ at a given position $x_i$ in the trap.

To model the Ramsey fringe, we assume that the atoms are in a two-dimensional harmonic trap with a radial distribution $P(r) = \frac{1}{\pi\rho w^2}e^{-r^2/\rho w^2}$, where $w = 16$ $\mu$m is the waist of the cavity mode for the 780-nm drive field, and $\rho$ is the local ratio of temperature to trap depth. We parameterize the vector light shift $\Omega_i(r) \approx (-1 + 2r^2/w^2)\Omega_i$ in terms of the magnitude $\Omega_i$ of the light shift on cavity axis ($r = 0$) for an atom in a lattice site at position $x_i$, averaged over the fast axial motion within the lattice site. We expect the magnetization at position $x_i$ to evolve as

$$\frac{f_i^z}{f_i}(t) = \int_0^{\Omega_i} P(\Omega_i(r))e^{-i\bar{n}\Omega_i(r)t}d\Omega_i(r) = \int_0^1 \frac{1}{2\rho\Phi_i}\Phi_i^{1/2\rho}e^{-i\bar{n}\Phi_i(\tau-\tau_0)}d\Phi_i, \quad (S39)$$

where $\tau = \Omega_i t$, $\tau_0$ allows for an initial phase, and $0 \le \Phi_i \equiv \Omega_i(r)/\Omega_i \le 1$ is the local probe coupling in units of $\Omega_i$. Fitting Eq. S39 to the data at $x_i$ yields a value for $\bar{n}\Omega_i$, and we divide by the intracavity photon number $\bar{n}$ to obtain $\Omega(x_i) \equiv \Omega_i$.

### D. Extracting the Interaction Strength from Flip-Flop Measurements

To measure the strength and sign of the couplings $\chi_{ij}$, we first use a pair of Raman beams to prepare a region (A) of the atoms along $\hat{F}_x$ and another region B along $\hat{F}_y$. In particular, after initializing all atoms in the $|m_f = -1\rangle$ state, we use a focused Raman beam to apply a local $\pi/2$ rotation to only the atoms in region (A). A second Raman

beam, phase-shifted by $\pi/2$ relative to the first beam, then addresses all atoms to apply a global $\pi/2$ rotation to the entire cloud. We then turn on photon-mediated interactions. In addition to coherent four-photon processes that cause each spin to precess about the mean field of the other spins, dissipative two-photon processes cause the spins to drift toward $m_f = \pm 1$, leading to a bias in the total magnetization. We extract both the mean interaction rate and mean decay rate by analyzing the sums and differences of the magnetization signals in the (A) and (B) regions.

To analyze the mean-field flip-flop dynamics, we first write the Hamiltonian and Lindblad operators of Eqs. S21 and S23 explicitly in terms of the local magnetization $\mathbf{F}_n \equiv \mathbf{F}(x_n)$:

$$H = 2\chi \sum_{n,m} \xi_n \xi_m \left( F_n^x F_m^x + F_n^y F_m^y \right) + \sum_n h_n F_n^z, \tag{S40a}$$

$$L_\pm = \sqrt{\gamma_\pm} \sum_n \xi_n F_n^\pm. \tag{S40b}$$

The indices $n, m$ label discrete regions in the 1-dimensional image obtained after summing over the transverse dimension of the cloud. Each region $n$ corresponds to a sum over many individual atoms, e.g.:

$$\mathbf{F}_n = \sum_{i \in n} \mathbf{f}_i, \tag{S41}$$

where we have assumed that the cavity couplings $\Omega(x)$ are approximately constant over the sites within each region. This coarse-graining over individual atoms justifies a classical calculation of the time evolution, where we approximate the $10^3 - 10^4$ uniformly polarized quantum spins in each analysis region as a single classical spin and neglect quantum correlations.

At short times, the classical spin model predicts the average magnetization $F_n^z$ in region $n$ to grow linearly in time at a rate dependent on the initial spin texture:

$$\frac{d}{dt} F_n^z = 2\xi_n \left( \mathcal{F}^x F_n^y - \mathcal{F}^y F_n^x \right) \chi - \xi_n \left( \mathcal{F}^x F_n^x + \mathcal{F}^y F_n^y \right) \gamma \tag{S42}$$

where $\gamma = \gamma_- - \gamma_+$ and $\mathcal{F} = \sum_n \xi_n \mathbf{F}_n$ is the collective spin, weighted by the cavity couplings. By measuring the linear growth rates in two different regions $n, m$ we obtain a pair of linear equations:

$$\frac{d}{dt} \begin{bmatrix} F_n^z \\ F_m^z \end{bmatrix} = \mathcal{M} \begin{bmatrix} \chi \\ \gamma \end{bmatrix} \tag{S43}$$

where the matrix $\mathcal{M}$ depends on the initial orientations of the spins in the two regions. We obtain an estimate for the initial local spin polarization $\mathbf{F}_n$ using the known spatial dependence of the two Raman beams, and we use the measured vector light shift $\Omega(x)$ (main text, Fig. 2b) to estimate the weights $\xi_n$. Inverting the matrix $\mathcal{M}$ then gives an estimate of the interaction and decay rates $\chi, \gamma$.

---



\end{document}